# Time-Domain Classification of the Brain Reward System: Analysis of Natural and Drug Reward Driven Local Field Potential Signals in Hippocampus and Nucleus Accumbens

**Amirali Kalbasi[a], Shole Jamali[b], Mahdi Aliyari Shoorehdeli[a,*], Alireza Behzadnia[c], Abbas Haghparast[d]**

[a] *Department of Mechatronics, Faculty of Electrical Engineering, K. N. Toosi University of Technology, Tehran, Iran*

[b] *Department of Neuroscience, Medical University of South Carolina, Charleston, SC, USA*

[c] *Histopathology Department, Leeds Teaching Hospital NHS Trust, Beckett Street, Leeds, West Yorkshire, LS9 7TF*

[d] *Neuroscience Research Center, School of Medicine, Shahid Beheshti University of Medical Sciences, Tehran, Iran*

***Numbers of pages: 34 pages***

***Number of figures: 7 figures***

* **Correspondence should be sent to:**

Mahdi Aliyari Shoorehdeli, PhD

Department of Mechatronics, Faculty of Electrical Engineering

K. N. Toosi University of Technology

**Tel.:** +98(21)84062-403

**Fax:** +98(21)84062-172

**P.O.Box**: 16315-1355, Tehran, IRAN

**Email:** Aliyari@kntu.ac.ir

## Abstract

This study investigates the neural underpinnings of substance use disorder (SUD) by employing novel time-domain analysis of Local Field Potentials (LFP) to classify and differentiate the neural pathways activated by natural- (food) versus drug-induced (morphine) rewards within the hippocampus (HIP) and nucleus accumbens (NAc) as critical regions in the brain reward system. We employ a robust preprocessing and validation framework to ensure signal integrity using autocorrelation and chaos-detection techniques, including Lyapunov and Hurst exponents. Subsequently, classification leverages the divergence in probability density functions (PDF) of LFP signals from the HIP and NAc using Jensen–Shannon divergence (JSD) to classify reward sources including food, morphine or saline (control). Interestingly, the Symmetrized Dot Pattern (SDP) technique is used for visual differentiation and categorization of signals from the HIP and NAc regions. The proposed method effectively classifies and differentiates neural activities within the HIP and NAc regions, based on the origin of the reward stimulus. This method represents significant advancement in real-time neural signal analysis of deep brain activities relevant to understanding neurobehavioral conditions like reward related behaviors, offering a remarkably fast and computationally light alternative to traditional and deep learning techniques. Its unique ability to validate, visualize, and categorize neural signals through its PDF and SDP techniques not only sets it apart from existing methods but also ensures interpretability and ease of use in clinical settings, even with limited computational resources.

## 1. Introduction

Signal processing methods are essential tools for understanding and interpreting neural circuit activities. The challenge is choosing the most appropriate analysis method since no universally accepted method exists which has significant implications for the insights derived from the data [1-3]. This manuscript emphasizes the potential of time-domain analysis, offering a more interpretable and less computationally intensive alternative to frequency and time-frequency domain methods [4-6]. Unlike supervised or unsupervised learning techniques, which may obscure biological mechanisms behind black-box operations [7, 8], our approach aims to preserve and highlight these features, ensuring a clearer understanding of neural signals.

Studies of neural circuits in the literature often focus on frequency or time-frequency domains of signals, with only a few employing time-domain signal analysis [9]. Time-domain analysis offers several advantages that make it well-suited for real-time applications, including its inherent simplicity, ease of interpretation, and lower computational requirements compared to alternative methods. [10, 11]. While Prior work on time-domain feature extraction primarily utilized basic signal characteristics—mean, variance, skewness, and others—to classify electroencephalogram (EEG) signals for detecting epileptic seizures [12, 13], our study innovates by applying more advanced time-domain analysis to classify local field potentials (LFP) signals in compulsive reward-seeking behavior, focusing on drug- vs. natural-rewards. This represents a novel application in a field where understanding the neural basis of addiction is crucial.

Addiction is a major public health concern characterized by frequent and compulsive reward-seeking behavior, in the form of unnatural substance abuse or natural forms of addiction, including food addiction, hypersexuality, or gambling. When considering natural and unnatural sources of reward, although some studies have suggested behavioral and neurochemical differences between the two, limited studies have compared the neural pathways activated via drugs and naturally induced types of rewards [14-17]. This gap highlights the uniqueness of our approach, applying time-domain analysis to explore the neural underpinnings of SUD, focusing on the differentiation between drug-related and natural rewards. By delving into these specific neural circuits, our study aims to investigate the complex mechanisms of SUD, contributing to a deeper understanding of its neural basis [18].

Within the intricate network of neural circuits underlying SUD, the mesolimbic pathway stands as the cornerstone of the reward system, with the ventral striatum (VS) and ventral tegmental area (VTA) being the most important [19]. The effector neural projections from the VTA to the nucleus accumbens (NAc) are associated with motivation and reward whilst the excitatory glutamatergic signals from the hippocampus (HIP) to NAc mediate learned behavior in addiction [20]. Hippocampal activity is also important in reward acquisition and expression phases [21-23]. In this study in order to investigate the functional connectivity of NAc and HIP as two crucial areas of a reward system in drug vs. natural sources of reward, a conditioned place preference paradigm (CPP) was designed [24] using saline as a control group, morphine and food to study the drug vs. naturally induced LFP in NAc and HIP simultaneously, in male Wistar rats' brain, by considering a novel approach in analyzing the time-domain feature of the LFP signals [25, 26].

In our approach, the validation of time-domain neural data leverages both nonchaotic and chaotic methodologies, including autocorrelation, Hurst, and Lyapunov exponents. This meticulous validation ensures the dataset's richness and applicability for further analysis. Subsequently, the classification of data into drug vs. natural reward sources utilizes Jensen–Shannon divergence (JSD) to differentiate morphine and food-driven pathways by analyzing probability density functions (PDF) of LFP signals in the HIP, NAc, and their interactions. JSD quantifies the information variance between PDF, highlighting distinct signal pathways. The symmetrized dot pattern (SDP) technique visually differentiates HIP and NAc signals, enabling rapid identification of amplitude variations. This study pioneers the separation and classification of LFP time-domain signals in deep and focal brain regions through JSD and SDP, marking a novel approach to neural circuit preprocessing.

The proposed approach successfully classified LFP signals into a source of reward (morphine or food, and saline as control) and the source of signal from the brain area (HIP vs. NAc) which is validated through statistical tests and visual patterns, achieving 100% accuracy in classification tasks. This high level of precision underscores the potential of our approach to provide meaningful insights into deep neural activities, particularly in the context of addiction research. Such validation highlights the computational efficiency and real-time applicability of our method in both laboratory and clinical settings, offering a significant advancement in the field of neuroscience.

Our study contributes to the body of knowledge by integrating engineering and data analysis perspectives into neuroscience research. It further highlights the role of time-domain signal analysis in understanding neural circuit activities, bridging the gap between technical methods and their applications in biological mechanisms.

The content of this paper was structured as follows: The behavioral and electrophysiological tests are presented in Section 2, Section 3 explains the data hierarchy and data preprocessing and validation. Section 4 describes the time-domain classification methods and results, and Section 5 summarizes and concludes the study.

## 2. Behavioral and electrophysiological tests

### *2.1. Animals and Surgery*

Male Wistar rats (Pasteur Institute, Tehran, Iran; weighing 220–270 g) were housed under standard laboratory conditions (12/12h light/dark cycle in temperature (25±2°C) and humidity (55±10%). Animals were restricted for food until they were reduced to 80-85% of their free-feeding body weight before the conditioning phase. The Ethics Committee approved all protocols of Shahid Beheshti University of Medical Sciences, Tehran, Iran (IR.SBMU.SM.REC.1395.373) and followed the National Institutes of Health Guide for the Care and Use of Laboratory Animals (NIH Publication; 8th edition, revised 2011). The rats were anesthetized using a cocktail of Ketamine and Xylazine (100/10 mg/kg; intraperitoneal (IP)) and placed stereotaxic apparatus (SR-8N, Narishige, Japan); the skin was incised, the skull was cleaned, and a hole was opened above of each target areas hippocampal CA1 (anteroposterior: -3.4 mm, lateral: ±2.5 mm); NAc (anteroposterior: +1.5 mm, lateral: ±1.5 mm). The recording electrodes were implanted into HIP (dorso-lateral: -2.6 mm) and NAc (dorso-lateral: -7.6 mm) to simultaneously record local field potentials (LFPs) in these areas. The reference and ground screws were inserted in the skull (Fig. 1.A.a-d). The rats were allowed to recover for ten days following surgery [27-29].

### *2.2. Drugs and Food*

Ketamine and Xylazine were obtained from Alfasan Chemical Co, Woerden, Holland. Morphine sulfate (Temad, Iran) was dissolved in physiological saline (0.9% NaCl) and administered by a subcutaneous (s.c.) route at the dose of 5mg/kg for morphine (drug) conditioning and 6g of Hobby biscuits for food conditioning.

*2.3. Conditioned place preference paradigm*

The conditioned place preference paradigm (CPP) apparatus consisted of three-chamber polyvinyl chloride boxes; two large side chambers (equal size) were connected to a smaller one (Null; with a smooth PVC floor). The two larger chambers differed in their floor texture (smooth or rough) and wall stripes pattern (horizontal or vertical) that provided distinct contexts that were paired with morphine, food, or vehicle (saline injections). Guillotine doors separated three distinct chambers. The CPP procedure consists of three phases: 1) pre-conditioning, 2) conditioning, and 3) post-conditioning.

During the pre-and post-conditioning phases, rats freely explored the entire arena for 10 minutes while they were connected to the LFP recording cable.

1) **Pre-conditioning phase**

Twenty-four hours before the conditioning phase, each animal was given free access to all three compartments. The animal was placed in the CPP apparatus for 10min. The time spent in each compartment was recorded and measured using a 3CCD camera. Animals that showed an inherent preference, defined as spending more than 80% of the time in one compartment during CPP, were excluded from the experiment (Fig. 1.B.a).

2) **Conditioning phase (Saline, Morphine, Food)**

For morphine conditioning, each animal received the 5 mg/kg of morphine (s.c.) as a drug reward and was placed and restricted (the sliding door was kept closed) in the reward-paired compartment (rewarded) of the CPP apparatus for 30min. For food conditioning, animals received food pellets (6g of Hobby biscuits) in the middle of the reward-paired compartment (rewarded) for 30min. Following six h, each animal in the drug- or food-CPP group experienced an injection of saline (1 ml/kg; s.c) as the vehicle of morphine, or nothing, respectively, in the non-reward compartment (unrewarded) for 30min. On the following day, the previous day, the morning protocol was switched to the afternoon and vice versa throughout the experiment. The mentioned protocol was conducted to the end of the conditioning phase (three days for the morphine CPP and five days for the food CPP) (Fig. 1.B.b) [30, 31].

3) **Post-conditioning phase**

Twenty-four hours after the conditioning phase, each animal was tested under a food- or morphine-free condition with free access to all three compartments. The animal was placed in the CPP apparatus for 10min. The time spent in each compartment was recorded and measured using a 3CCD camera. The conditioning score (CPP score) was calculated by subtracting the time spent in the unrewarded paired compartments from the rewarded paired compartment (Fig. 1.B.c).

*2.4. Behavioral and LFP recording*

LFP recordings were collected from hippocampal CA1 and NAc of freely moving rats during pre-and post-conditioning while the animal performed the CPP procedure. The animal's behavior was recorded with a digital video camera (30 frames per second), and movements were tracked by an automated system synchronized with behavioral data and electrophysiological recordings. The spatial position was defined as the animal's head in each frame. A lightweight and flexible cable was connected to the pins on the head-stage preamplifier. Recordings, digitalization, and filtering of neural activities were performed using a commercial acquisition processor (Niktek, IR). LFP recordings were sampled at 1000 Hz. Fig. 1.C.b shows a sample of recorded LFP from HIP (top) and NAc (bottom) for 10-min in the rat that had received morphine.

At the end of the experiments, the electrode trace was marked with the electrical lesion (25 μA, 10s) before the animals were perfused with isotonic saline followed by 10% formalin. Brains were sliced (150 μm) using a vibrating microtome (Campden Instruments, Germany). Electrode tip traces were localized using a light microscope and were confirmed using a rat brain atlas (Paxinos and Watson 2007).

## 3. Methodology

### 3.1. Data hierarchy and Signal Analysis Strategy

Data hierarchy was organized based on three aspects of the CPP experiment, namely the phase of the experiment, treatment received during the CPP conditioning period (morphine vs. food vs. saline), and the site-specific LFPs (HIP or NAc) (Fig. 1.C.a):

1. Conditioning-phase: pre-and post-conditioning phases correspond to the timing of LFP measurement before or after rats were conditioned to their treatment.
2. Treatment groups: animals received three different treatments during the conditioning phase: morphine, food, or saline.
3. Recording sites: local field potential signals recorded via implanted electrodes in HIP and NAc.

Data were analyzed by considering signals from either HIP or NAc separately (henceforth referred to as HIP/NAc) and the HIP and NAc connectivity (henceforth referred to as HIP-NAc) in the time domain, following three steps: 1) *preprocessing:* noise and outlier reduction; 2) *validation* to evaluate meaningfulness of the captured signals, and 3) *feature extractions and signal classification*. All analyses were performed using MATLAB 2020b (The MathWorks, Inc., US) and Python (v3.7).

*3.2. LFP signal preprocessing*

LFP signals were filtered within a 0.5 to 300 Hz frequency range to remove background noise and enhance the accuracy of neural activity recordings. Additionally, the three-sigma rule was applied to identify and correct outliers and saturated data points, which were adjusted by replacing them with the arithmetic mean. This step is crucial for ensuring the reliability and integrity of the LFP data, setting a solid foundation for precise downstream analysis and interpretation [32].

*3.3. LFP signal validation*

LFP recordings of neuronal activity within an area of interest may become contaminated by the activity of neurons outside the area of interest. The recordings can also get corrupted due to faulty hardware, leading to disrupted signals. To ensure the LFP recordings accurately reflect the neuronal activity within the target areas, a comprehensive validation process was implemented. This process includes assessing the signal for long-term dependence (i.e., reflective of the sequential structure of neural activities associated with specific behavior) [33-35] and chaotic characteristics (i.e., reflective of temporal dependency of neural activities on the initial excitatory or inhibitory signaling), crucial for distinguishing between genuine neural activities and potential artifacts or external interferences using 1- Hurst exponent (H), 2- autocorrelation and 3- Lyapunov exponent (LE) [36].

1- Hurst exponent is a measure of the long-term dependence and randomness of the time-series data [37]. Equation (1) shows the Hurst exponent equation where; $R(n)$ is the range of the first $n$ cumulative deviations from the mean, and $S(n)$ is their standard deviation $\mathbb{E}[x]$ is the expected value, -n is the period of the observation, $C$ is a constant [38].

H value of 0.5 indicates a lack of time-dependent correlation [39], whilst values higher or lower than 0.5 suggest positive or negative correlations between the values at $t_n$ in the series to the lagged values at $t_{n-1}$ respectively. In other words, the value of each data point (e.g., neural activity at $t_1$) is dependent on the preceding values in the series (e.g., neural activity at $t_0$) and is not purely random [40].

$$\mathbb{E}\left[\frac{R(n)}{S(n)}\right] = Cn^H \qquad (1)$$

2- Further, the Lyapunov exponent was calculated for all LFP signals originating from either HIP or NAc during pre-conditioning and post-conditioning phases, regardless of their treatment group (Fig 2.A), to identify the presence of chaotic dynamics. Equations (2) describe the Lyapunov exponent ($\lambda$), which quantifies LFP signals sensitivity to initial conditions. In this context, $\delta\boldsymbol{Z}(t)$ refers to the difference between two state vectors of the LFP signals. Each state vector represents the condition of the signal—such as its amplitude and rate of change—at a specific moment in time. These vectors evolve as the signal progresses and are plotted in phase space, a conceptual space used to visualize the signal's dynamics (e.g., signal amplitude and its derivative) over time. The Lyapunov exponent measures how quickly these initially close trajectories, separated by $\delta\boldsymbol{Z}(t)$, diverge as time progresses [41, 42].

$$\lambda = \lim_{t\to\infty} \lim_{|\delta\boldsymbol{Z}_0|\to 0} \frac{1}{t} \ln \frac{|\delta\boldsymbol{Z}(t)|}{|\delta\boldsymbol{Z}_0|} \qquad (2)$$

Equation (3) shows that this divergence grows exponentially over time, illustrating that even a small difference in initial conditions can rapidly increase, indicating chaotic behavior [43]. This sensitivity to initial conditions is crucial for distinguishing structured neural activity from potential noise [44].

$$|\delta \boldsymbol{Z}(t)| \approx e^{\lambda t}|\delta \boldsymbol{Z}_0| \tag{3}$$

A positive $\lambda$ close to zero indicates that the neural activity is indeed sensitive to initial conditions, allowing for a clear distinction between structured neural activities and noise [45]. Large maximal Lyapunov exponents, on the other hand, are characteristic of irregular and non-periodic chaotic signals (similar to white noise or impact noise) [46].

3- Autocorrelation which is the correlation of a signal with its delayed self [47], measures provide an additional layer of validation, confirming the signal's internal consistency by highlighting the relationship between sequential data points [48]. A non-zero autocorrelation in the initial lags substantiates the signal's biological origin and its relevance to our study [49]. Equation (4) shows autocorrelation formula where $R_{XX}$ represents the autocorrelation function, which measures the correlation between the values of the same random process X(t) at two different times, $t_1$ and , $t_2$. $E$ denotes the expected value and $\overline{\mathrm{X}}$ is the complex conjugate of the value of the random process X(t).

$$R_{XX}(t_1, t_2) = E\left[X_{t_1}\bar{X}_{t_2}\right] \tag{4}$$

Validation tests conducted on the LFP signals recorded from both NAc and HIP of all subjects revealed that the Hurst exponent of the data was significantly elevated (H ≈ 1) when compared to randomly shuffled LFP signals (H ≈ 0.5). These comparisons were performed using 5-second, non-overlapping windows (Fig. 2B), therefore, the recorded

LFP signals are meaningful, which indicates a predominant excitatory sequential neural signaling pattern. Additionally, the recorded LFP signals' autocorrelation was non-zero (≈ 0.95 in the first lag) (Fig 2.C). Furthermore, the global Lyapunov exponent of all rats shows a positive rate, close to zero that can be interpreted as a weakly chaotic behavior capable of short-term predictions [50]. The validation process, leveraging the Hurst exponent, Lyapunov exponent, and autocorrelation measures, confirms the reliability of the LFP recordings. High Hurst values and near zero Lyapunov exponent outcomes, complemented by significant autocorrelation, indicate that the data are well-suited for subsequent analyses aimed at understanding neural mechanisms of addiction [51, 52].

### *3.4. Hypothesis-test for validation of classification methods*

Hypothesis testing is used to validate classification methods by identifying significant differences between categorized groups. First, the Kolmogorov-Smirnov test evaluates normality. For non-normally distributed data, the Wilcoxon rank test and Mann-Whitney U tests used to measure significant differences within and between groups. Conversely, for normally distributed data, t-tests (paired and unpaired) assess two-group differences, while ANOVA, followed by Tukey's post-hoc test, used for multi-group comparisons, maintaining an alpha threshold of 0.05.

## 4. Time-domain-based features classification

### *4.1. Basic features*

The initial analysis consisted of separating and classifying LFP signals into three groups (morphine, food, or saline); two phases (pre-and post-conditioning); and two recording sites (HIP and NAc) using basic and common characteristics such as mean, maximum, minimum, median, and standard deviation. However, the LFP signals did not appear to be significantly separated into the desired categories; therefore, other features of the signals were considered.

*4.2. HIP-NAc correlation*

The LFP signals were classified by considering the functional connectivity between the HIP and NAc regions. The Pearson correlation - chosen for its ability to quantify the linear relationship between these brain regions' activities- of HIP-NAc signals shows a strong significant correlation between these two regions in post-conditioned subjects being treated with food, while the other treatment groups did not show a significant or meaningful correlation ($-0.1 < r < 0.1$) (Fig 3). No correlation was found among the three treatment groups in the pre-conditioning phase.

These results indicate that the functional connectivity between HIP-NAc was increased following food treatment as a natural reward while it did not change following morphine reward; therefore, HIP-NAc functional connectivity is required for inducing food reward but not for the morphine reward induction.

*4.3. One-dimensional probability distribution function*

The correlation analysis highlighted the functional connectivity between HIP and NAc; however, neuroscientists are interested in studying the signal changes in each specific site separately. Given the LFP signals are continuous variables, the one-dimensional probability density function (1D-PDF) can be assumed to classify HIP and NAc according to the type of reward received based on the Bayes theorem. Uncovering the underlying PDF of data also enables us to have a more comprehensive description of the signal.

The maximum likelihood estimation method (MLE) estimated the most probable underlying density function [53] by considering multiple distributions. Results show that the Gaussian distribution was the most probable PDF of the LFP signals. The PDF of a Gaussian distribution ($f(x)$) is defined using two parameters, the mean ($\mu$) and variance ($\sigma$) (5) [54].

$$f(x) = \frac{1}{\sigma\sqrt{2\pi}} e^{-\frac{1}{2}\left(\frac{x-\mu}{\sigma}\right)^2} \quad (5)$$

The Gaussian PDF estimated parameters of LFP signals from HIP (Fig. 4.A; left panel) and NAc (Fig. 4.A; right panel) regions were compared between pre-and post-conditioning phases in each saline, morphine, and food treatment groups (Fig. 4.A).

The results showed a significant difference in the PDF variance of the LFP signals in the NAc region of the subject treated with morphine (Fig. 4.A middle-right panel); with a reduced and narrower variance and amplitude of LFP signaling in post-conditioning compared to pre-conditioning. This reduction in the signaling amplitude has also been reported previously, where a reduced NAc signaling was observed following chronic morphine treatment [55]. There was no difference in NAc activity of the other two treatment groups between pre- and post-conditioning (Fig. 4.A top- and bottom-right panels).

Interestingly, the reverse in the HIP region in subjects treated with food (Fig. 4.A bottom-left panel), a variance of the HIP PDF after food-CPP was increased. No significant difference was seen in other groups (Fig. 4.A top- and middle-left panels).

According to these results, the conditioning phases could be separated based on the changes in the PDF parameters. Additionally, treatment groups could be separated. This was followed by Kullback–Leibler divergence to examine the difference between the PDFs of LFP signal more precisely.

The Kullback–Leibler divergence (KLD) was used to measure the similarity of LFP signals PDFs in pre-and post-conditioning phases of each treatment group in both HIP and NAc. The KLD score is widely used to measure the similarity and relative entropy between two distributions [56, 57].The KLD aims to measure the information gained between two PDFs. For two probability distributions (P and Q) in the same domain C, the KLD of P from Q is defined as (6) [58].

$$KLD(P \parallel Q) = \sum_{c \in C} p(c) \, log_2 \frac{P(c)}{Q(c)} \qquad (6)$$

The divergence is not symmetric. Thus $KLD(P \parallel Q) \neq KLD(Q \parallel P)$, with a non-negative possible value ($0 \leq$ KLD$< +\infty$) [59]. The KLD is 0 if the two distributions are equal in their outcomes, namely $P(c) = Q(c), \forall c \in C$. It has no upper-bound, as shown by Gibbs' inequality [60]. Jensen–Shannon divergence (JSD) is used to solve this problem. The JS divergence is defined as (7) [61].

$$JSD(P, Q) = \frac{KL(P \parallel A) + KL(Q \parallel A)}{2} \qquad (7)$$

For measuring JSD test, custom code was used in python 3.7.

To classify treatment groups into saline, morphine and food, the PDF similarity of pre- and post-conditioning phases were computed by JS-test. The higher calculated JSD indicates a more significant difference between PDFs (or less similarity).

Figure 4.B left panel shows that JSD of the HIP was significantly higher in the food-CPP group compared to saline (One-way ANOVA test with post-*hoc* Tukey test, $P<0.01$) and morphine treatment groups (One-way ANOVA test with post-*hoc* Tukey test, $P<0.01$);

there was no significant difference between JSD of saline and morphine groups. Figure 4.B right panel shows that the JSD of NAc in the morphine group was more than both saline (One-way ANOVA test with post-*hoc* Tukey test, $P<0.05$) and food (One-way ANOVA test with post-*hoc* Tukey test, $P<0.05$) treatment groups, while there was no difference between the JSD of NAc between saline and food group. These results show that we can classify the LFP signals based on the calculated JSD for each recording site (HIP and NAc) into either food or morphine-CPP.

To follow the goal of classifying each animal's recorded LFP signals into their respective treatment group, a threshold was calculated for the JSD score based on the LFP recording of the saline group (control group) as the baseline. Setting the HIP JSD threshold as >1000 (JSD threshold in the HIP for the saline group; HIP-JSD$_{T.saline}$), subjects treated with food were successfully classified as the food treatment group (accuracy = 100%), and NAc JSD threshold as >900 (JSD threshold in the NAc for the saline group; NAc-JSD$_{T.saline}$) classified the subject correctly as the morphine treatment group (accuracy = 100%). If the calculated HIP JSD was less than 1000 and NAc JSD was less than 900, subjects were classified as saline (accuracy = 100%).

These results suggest that, as expected, there is no change in HIP and NAc LFP signals when subjects were not conditioned in response to saline-CPP. HIP activity in animals that were conditioned with food and NAc in those conditioned with morphine has essentially changed their signaling patterns. Therefore, HIP activity plays an essential role in food-induced reward (natural reward) while NAc activity is necessary for morphine-induced reward (drug reward).

*4.4. Two-dimension multivariate probability distribution function (2D multivariate -PDF)*

A two-dimension multivariate probability density function of the connectivity of HIP and NAc (HIP-NAc 2D-PDF) was estimated to classify the different treatment groups. The most likely distribution for two-dimensional PDFs was identified using the previously described maximum likelihood estimation method (MLE). Gaussian 2D-PDF proved to be the best fit. x is an n×1 random vector. The density function for a multivariate Gaussian distribution ($p(x)$) with mean μ and covariance matrix is Σ [62]:

$$p(x) = \frac{1}{(2\pi)^{n/2} det\ (\Sigma)^{1/2}} exp\ \left(-\frac{1}{2}(x-\mu)^T \Sigma^{-1}(x-\mu)\right) \tag{8}$$

HIP-NAc 2D-PDF of LFP signals are shown in Fig. 5.A. The 2D Gaussian PDF properties (Covariance and Mean) were compared, and the results show that the covariance of the food treatment group had increased from the pre- to post-conditioning phase (one-sided t-test between pre-and post-conditioning, $P<0.05$).

To further investigate the HIP-NAc activity changes between pre-and post-conditioning phases, KLD of 2D Gaussian PDF was calculated [63]:

$$KLD(PDF_1 \parallel PDF_2) = \frac{1}{2}\left(log\frac{det\ \Sigma_2}{det\ \Sigma_1} - n + tr\ (\Sigma_2^{-1}\Sigma_1) + (\mu_2-\mu_1)^T \Sigma_2^{-1}(\mu_2-\mu_1)\right) \tag{9}$$

Where 1 and 2 refer to pre-and post-conditioning, respectively.

2D-HIP-NAC JSD was calculated and compared across all treatment groups to classify treatment groups. Results show that the food treatment group had the higher HIP-NAC

JSD, followed by morphine (one-way ANOVA test with post-*hoc* Tukey test, $P<0.01$), and lastly, saline groups (one-way ANOVA test with post-*hoc* Tukey test, $P<0.05$). Also, results show that the morphine treatment group had a higher HIP-NAc JSD than the saline treatment group (one-way ANOVA test with post-*hoc* Tukey test, $P<0.01$) (Fig. 5.B). This step-wide rise enabled us to set a 2D HIP-NAc threshold, to classify subjects with 100% accuracy into the food treatment group (0.3050 < Food-threshold), morphine (0.1341 < Morphine-threshold < 0.305), and saline group (Saline-threshold < 0.1341).

*4.5. Symmetrized dot pattern analysis (SDP)*

This part used the symmetrized dot pattern technique (Fig.6) for real-time classification and validation of recording LFP signals from deep brain regions.

The SDP method is used to visualize changes in the amplitude of time-series data that can be quickly interpreted and transform the time waveform into a scatter plot space. SDP visualizes data based on plotting each data point with bilateral mirror symmetry, having a specific six-fold symmetry shape [64]. The technique allows rapid detection of minor changes since symmetry is a visual property quickly and efficiently detected by human perception; therefore, minor changes in input signals will be noted [65]. SDP is defined as (10).

$$\begin{aligned} A(i) &= \frac{X(i)-X_{min}}{X_{max}-X_{min}} \\ \Theta(i) &= \theta + \frac{X(i+L)-X_{min}}{X_{max}-X_{min}}\zeta \\ \Phi(i) &= \theta - \frac{X(i+L)-X_{min}}{X_{max}-X_{min}}\zeta \end{aligned} \tag{10}$$

Where $X$ is the input signal, θ is the repetition or symmetry angle, L is a delay, ζ is the increase in the drawing angle, $A(i)$ are SDP radios, $\Theta(i)$ is the angle of a dot in SDP, $\Phi(i)$ is the negative angle of a dot in SDP, and $i$ is the number of selected points in signal ($i = t/\Delta t$) witch $\Delta t$ is the sampling rate. Experimentally, the parameters were selected as $L = 1, \theta = 45$, and $\zeta = 90$ [66].

Rapid plotting in SDP allows us to visualize the signals in real-time with many applications. For instance, this can be used to validate the placement of the LFP micro-electrodes for a specific region during a live experiment when an animal was performing the CPP as a behavioral task.

Interestingly SDP results of sample data of HIP (Fig.6, left panel) represent a pattern distinct from NAc visually, HIP is formed a vane pattern, whilst NAc is rather polygonal with a polygonal pattern (Fig.6, right panel).

## 5. Conclusion

This study aimed to validate, and classify LFP recorded data in the time-domain to investigate the neural activity of HIP, NAc, as well as connectivity of these two major regions of the reward circuit in response to morphine and food as drug and natural rewards, respectively. The methods used for data validation included chaotic tests (Hurst and Lyapunov tests) and a non-chaotic test (Autocorrelation). Various time-domain methods have been applied successfully for classification, with significant results from HIP-NAc correlation, differences in data PDFs, and SDP analysis for differentiating treatment groups, conditioning phases, and recording sites.

By employing the advanced methodologies outlined in this study (Fig.7), achieved 100% classification accuracy for LFP signals across several critical dimensions: reward type (food, morphine or saline (control)), signal origin (HIP or NAc), and conditioning phase (pre- vs. post-conditioning). This high-precision approach demonstrates not only the robustness of our analysis but also its practical application for real-time signal processing in both experimental and clinical environments. This study lays a solid foundation for future research, paving the way for further investigations into various neural circuits and their responses to different types of rewards using our methods. Our method provides a highly efficient, computationally streamlined solution that can be seamlessly integrated into real-time decision-making systems. These findings lay a crucial foundation for future research, facilitating more in-depth exploration of neural circuits and their involvement in reward processing. The ability to accurately classify and predict these variables paves the way for new insights into how the brain responds to different reward modalities and the underlying mechanisms governing reward-related behaviors.

Our study findings are particularly illuminating, revealing a clear differentiation in the neural signatures of natural versus drug rewards. Notably, our findings demonstrate that classification based on HIP data yields the best results for natural rewards, while NAc data are more indicative of morphine. Furthermore, our analysis revealed that morphine reduces HIP-NAc connectivity, whereas food appears to enhance this connectivity. These insights are pivotal, as they underscore the differential impact of natural versus drug rewards on neural circuitry, particularly in the context of the reward system connectivity which discussed in the next paragraphs:

### *5.1. HIP for Food Rewards*

*Role of the Hippocampus in Memory and Space:* The hippocampus plays a crucial role in memory formation and spatial navigation. Its involvement in processing natural rewards like food can be linked to the evolutionary importance of remembering the locations and contexts associated with food availability. This connection between spatial memory and food reward suggests why HIP data might yield the best results for natural rewards. The place cells allow hippocampus to create a cognitive map of the location animal received the food which is essential for navigating towards food sources.

*Contextual and Environmental Cues:* The hippocampus is sensitive to contextual and environmental cues that are often associated with natural rewards. This sensitivity is critical for survival, as it helps animals remember where and when food is available. The heightened activity in HIP in response to food rewards could reflect this intricate processing of contextual information [67].

*5.2. NAc for Morphine Rewards*

*Role of the Nucleus Accumbens in Reward and Pleasure:* The NAc is a key component of the brain's reward system, heavily involved in the processing of rewarding stimuli and pleasure. Its role is particularly pronounced in the context of drug rewards, where it mediates the acute effects of drugs of abuse, including opioids like morphine. *Dopamine Signaling:* Drug rewards often hijack the brain's natural reward pathways, with dopamine playing a central role. The NAc is a major target for dopamine, and morphine's ability to increase dopamine levels in this region could explain why NAc data are more indicative of drug rewards. This dopaminergic signaling is crucial for the reinforcing properties of drugs, making the NAc a focal point in studies of addiction [68].

*5.3. Differential Connectivity Between HIP and NAc*

*Impact on the Reward Circuit:* Findings on the differential impact of food and morphine on HIP-NAc connectivity are intriguing because they highlight how natural, and drug rewards can differently modulate the brain's reward circuitry. The enhancement of connectivity by food could be related to the need for integrating spatial and contextual information (HIP) with reward processing (NAc) to guide behavior towards survival-relevant goals. *Disruption by Morphine:* In contrast, morphine's reduction of HIP-NAc connectivity might reflect a disruption of this integration, leading to the prioritization of drug-seeking behavior at the expense of natural rewards and survival needs. This disruption can be a fundamental neurobiological substrate for the development of addiction, where the brain's natural reward systems are hijacked by drug rewards, leading to maladaptive behaviors [69].

These results provide crucial insights into the neural mechanisms underlying reward processing and how these pathways are differentially affected by natural versus drug rewards. By elucidating the specific roles of the HIP and NAc in response to different types of rewards and their connectivity, this study sets a precedent for future research to explore other neural circuits and their responses to rewards. It also offers potential pathways for investigating therapeutic strategies that could rebalance these reward systems in the context of addiction.

In conclusion, the specificity of HIP for food rewards and NAc for morphine rewards, along with the observed changes in connectivity, provide valuable insights into the brain's reward systems. These findings underscore the complexity of reward processing and its susceptibility to manipulation by drug rewards, offering important avenues for future research and therapeutic development.

## Figures and Legends

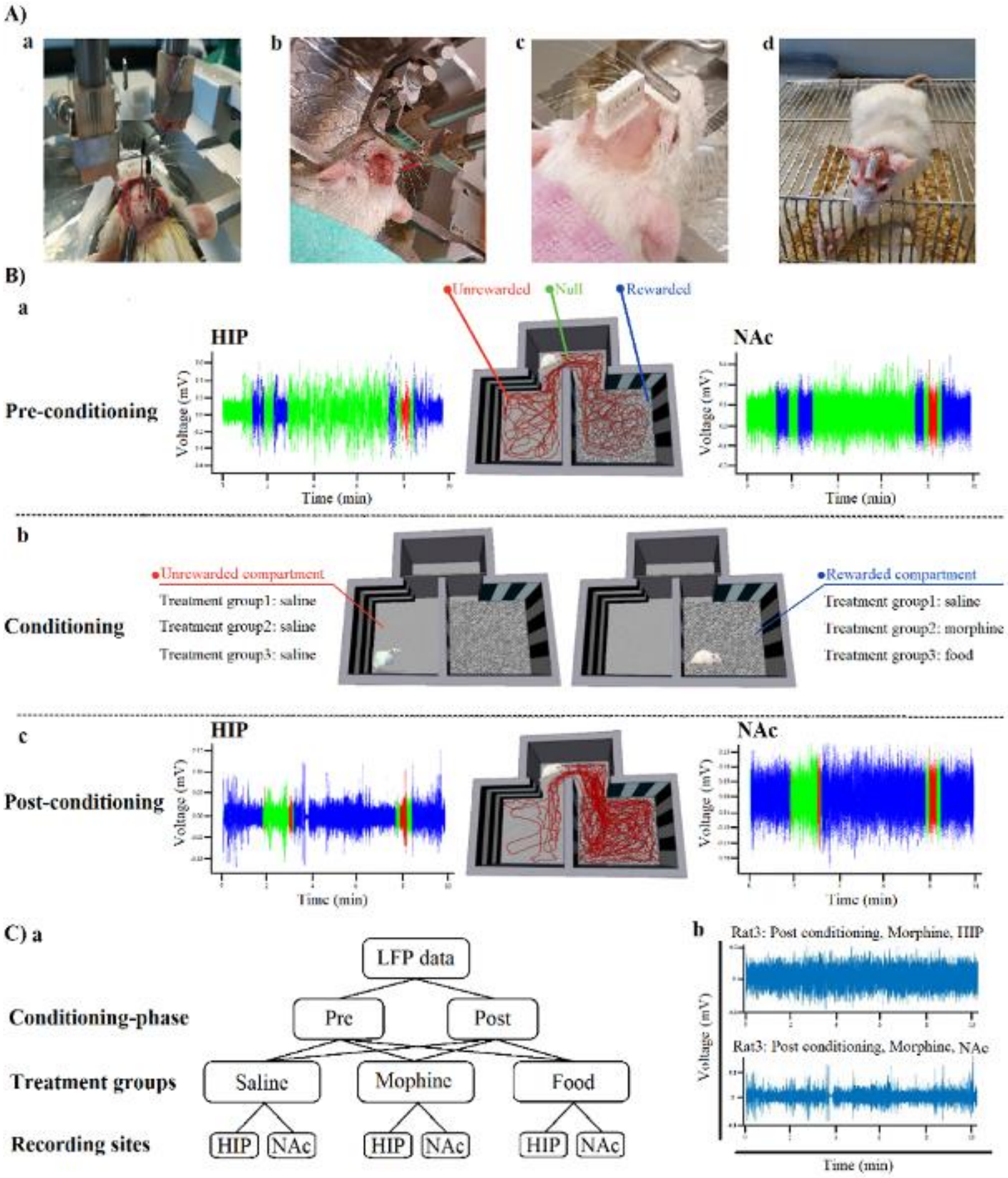

**Figure. 1.** Behavioral and LFP recording protocol. **(A.a-d)** Electrodes implementation surgery in HIP and NAc. **(B)** Conditioned place preference paradigm: **(B.a)** Pre-conditioning, a recorded sample from the HIP (left panel), schematic of freely moving rat (middle-panel), a recorded sample from NAc (right panel); duration of recording in both pre- and post-tests was 10 minutes; **(B.b)** Schematic of conditioning; in rewarded component animals received the rewards (morphine or food) or saline as control while they received the vehicles in unrewarded component. Duration of conditioning test was 30 minutes; **(B.c)** Post-conditioning: The left panel displays a recorded sample from the HIP, middle panel presents a schematic of a freely moving rat, demonstrating a preference for the rewarded component over the unrewarded component, right panel shows the LFP recording from NAc during the post conditioning. **(C.a)** Labeling structure, pre: pre-conditioning test, post: post-conditioning test; **(C.b)** Example of labeled data in one rat: The upper panel shows an example of LFP row data recorded from HIP during the 10-min post conditioning test for morphine, lower panel shows the recorded data from the NAc in the same session.

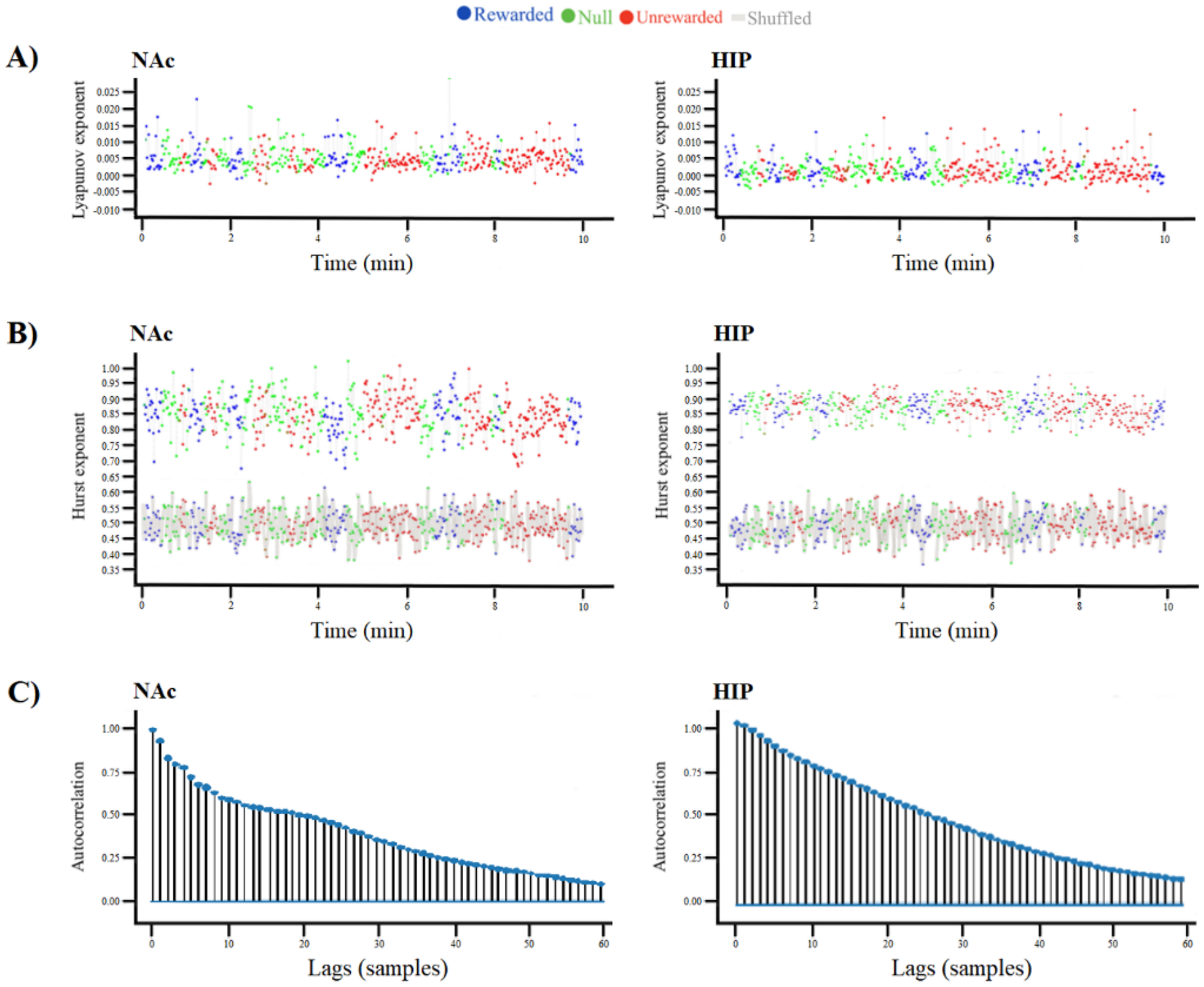

**Figure. 2. Validation tests. A)** Lyapunov exponent: The global Lyapunov exponent of LFP recording from NAc (left panel) and HIP (right panel) of all rats shows a positive rate, close to zero. The blue dots represent the Lyapunov exponent during the time spent in the rewarded compartment, green dots represent the null compartment, and red dots represent the unrewarded compartment. **(B)** Hurst exponent: The Hurst exponent of the raw LFP data was significantly elevated (H ≈ 1) for both NAc (left panel, upper) and HIP (right panel, upper) recordings compared to randomly shuffled LFP signals (H ≈ 0.5) (lower panels). **(C)** Autocorrelation: The left panel shows the non-zero autocorrelation in the initial lags substantiates for NAc LFP data, while the right panel represents the non-zero autocorrelation in the initial lags substantiates for HIP LFP data.

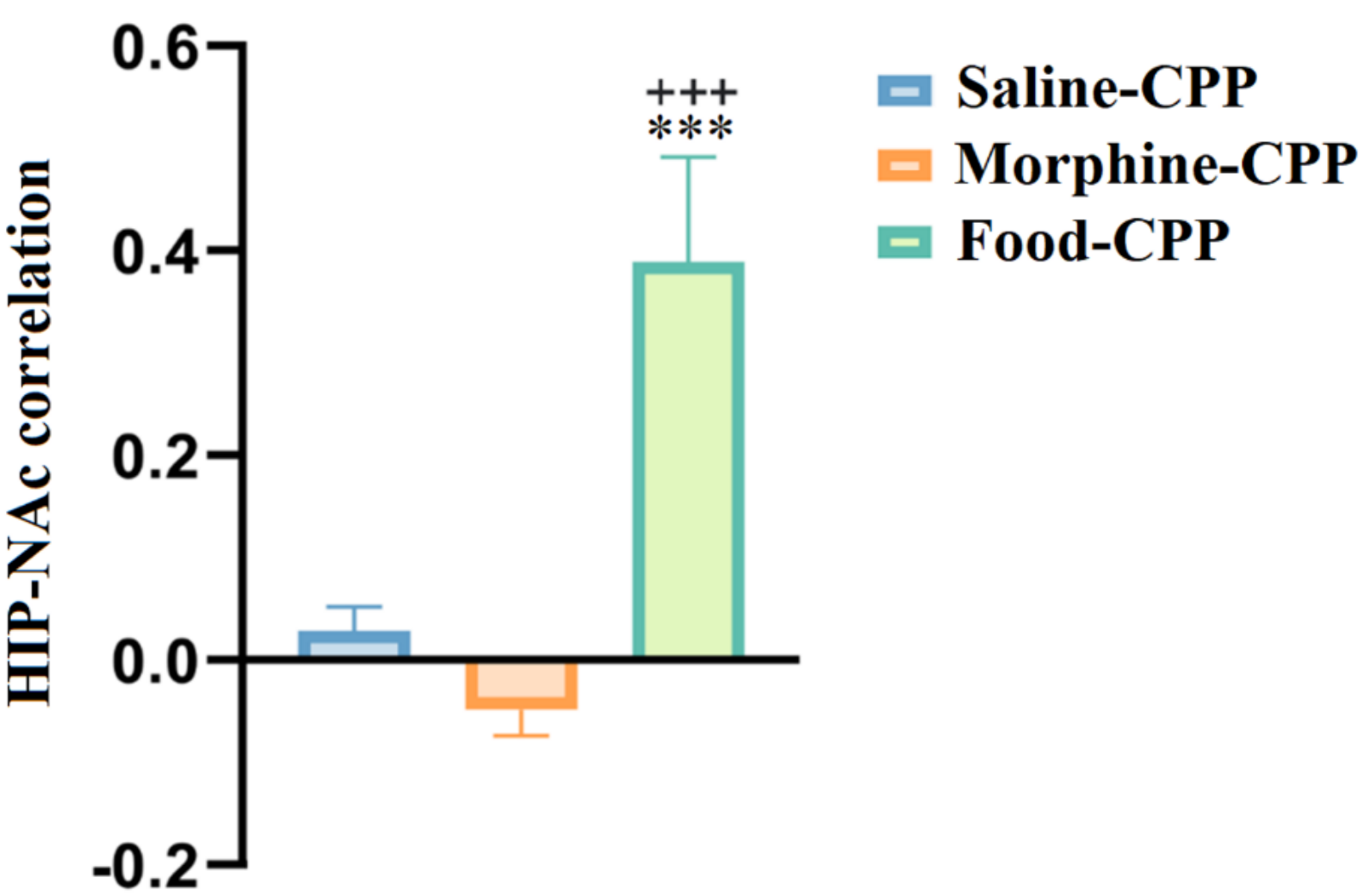

**Figure. 3. HIP-NAc correlation.** *** P<0.001 compared with the saline and morphine groups in the post-conditioning phase in CPP; shows an increase in HIP-NAc correlation in the food group compared to the morphine and saline groups, while a reversed correlation was observed in the morphine group.

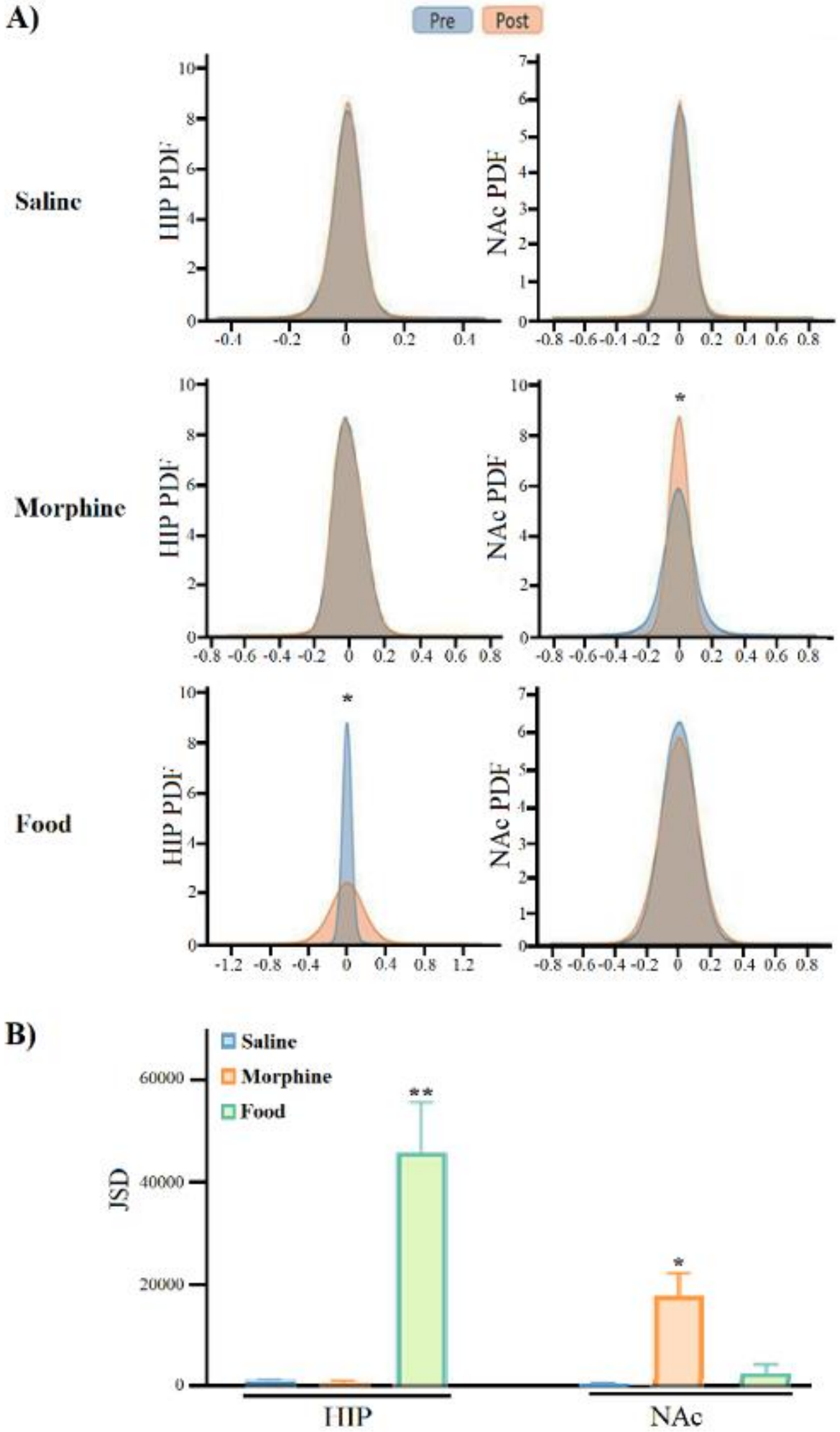

**Figure 1. One-dimensional probability density function (1D-PDF) of HIP/NAc. (A)** 1D-PDF of LFP recording data, *P<0.05 as a comparison between pre-and post-conditioning phases in each treatment's groups (saline, morphine, and food). **(B)** Comparing the JSD of 1D PDF of LFP signals among saline, morphine, and food groups,

**P<0.01 as a comparison of food group with saline and morphine groups in HIP; *P<0.05 as comparison of morphine group with saline and food groups in NAc.

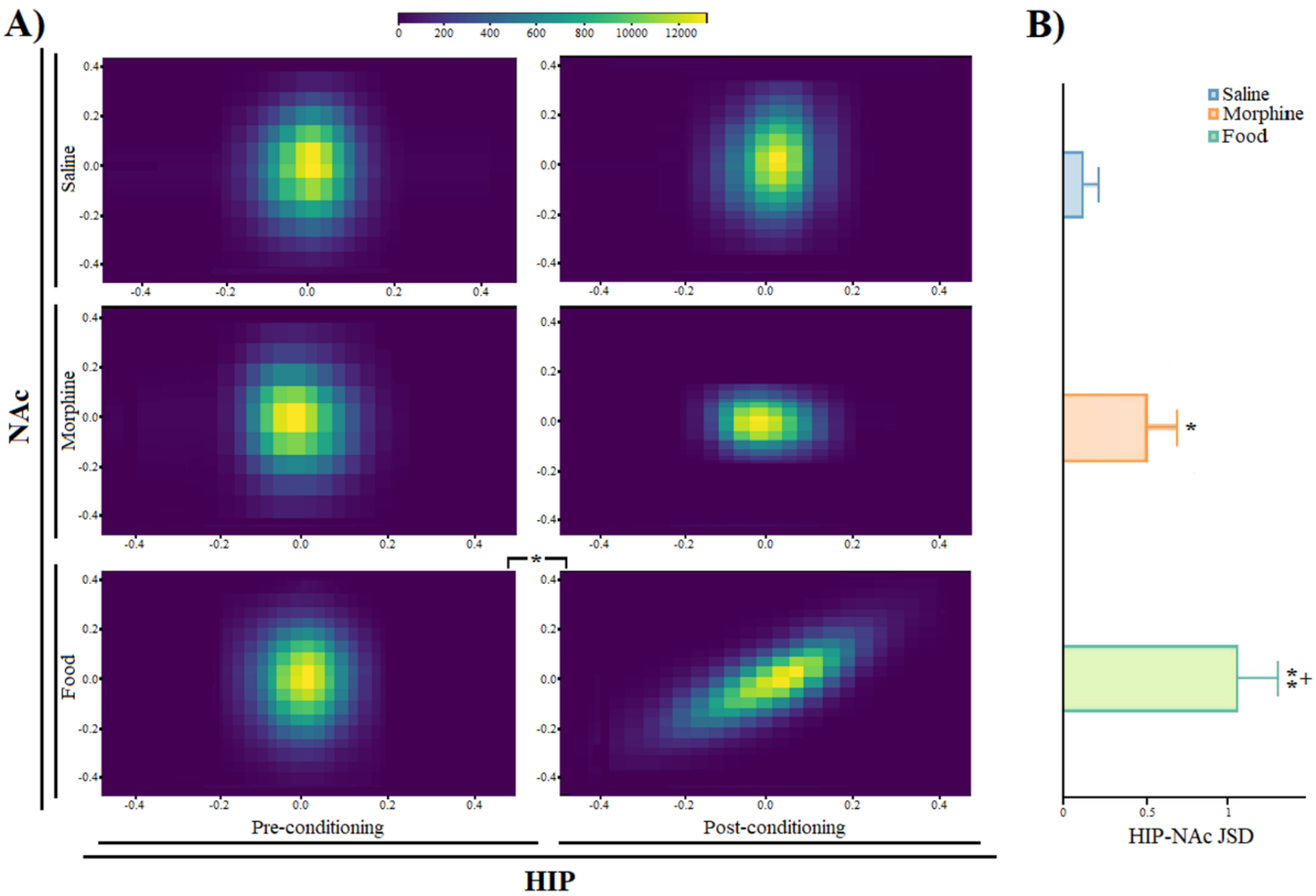

**Figure. 2. Two-Dimensional Probability Density Functions (PDFs) of Hippocampal (HIP) and Nucleus Accumbens (NAc) Activity.** (**A**) 2D HIP-NAc PDF histograms: HIP activity is plotted on the horizontal axis, and NAc activity on the vertical axis. The histograms display data for Saline (top row), Morphine (middle row), and Food (bottom row) treatment groups. For each treatment group, results are shown for pre-conditioning (left column) and post-conditioning (right column) phases. *P<0.05 denotes a significant difference between pre- and post-conditioning phases within the same treatment group. **(B)**

Jensen-Shannon Divergence (JSD) of 2D HIP-NAc PDFs: The JSD values quantify differences in the 2D PDFs between treatment groups. *P<0.05 as a comparison between saline and morphine groups, **P<0.01 as a comparison between saline and food groups, +P<0.05 as a comparison between morphine and food groups.

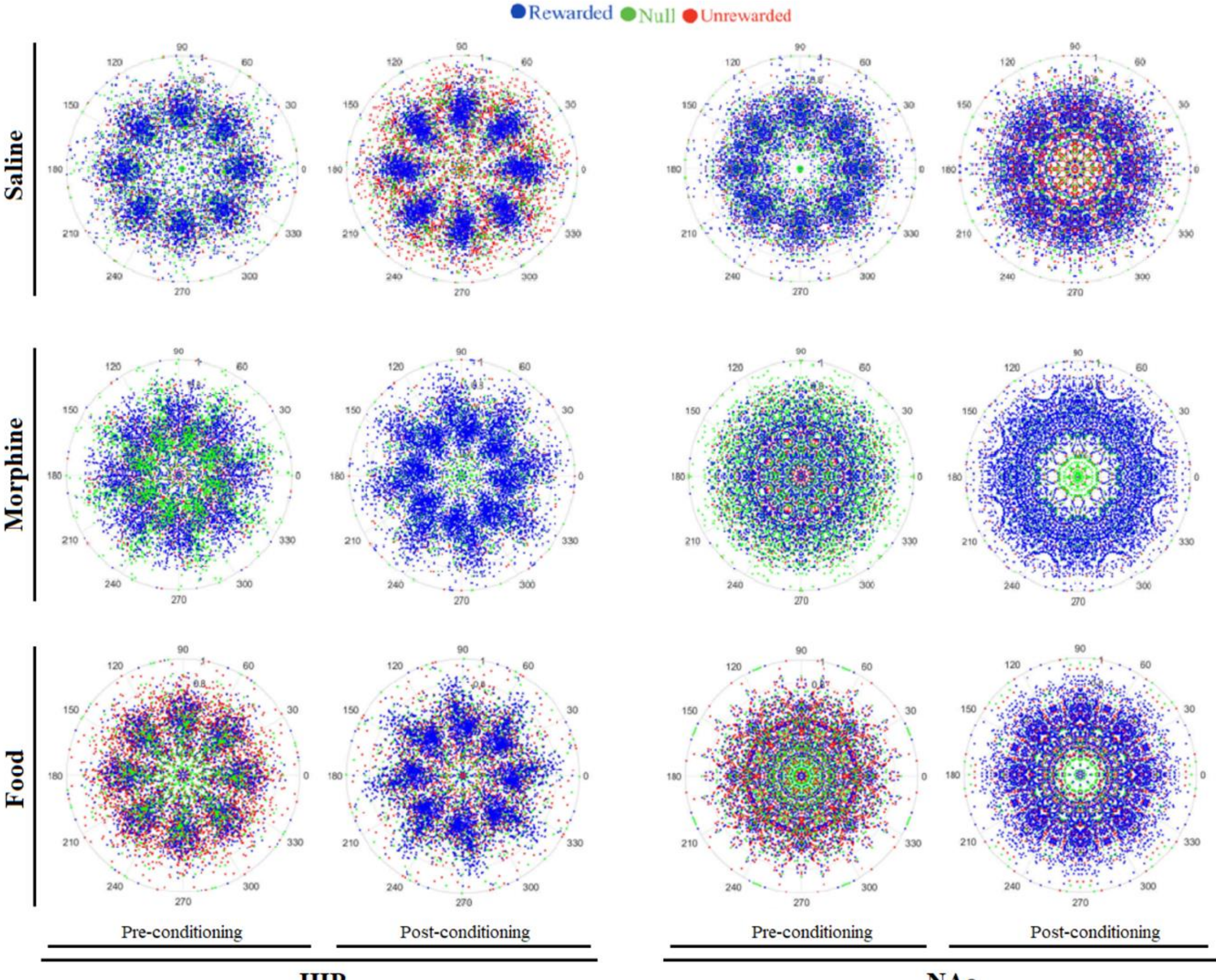

**Figure. 3. Symmetrized dot pattern (SDP) visualization.** Top panel: saline, middle panel: morphine, and bottom panel: food treatment groups; left panel: HIP, and right panel NAc recording sites; first and third columns: Pre-conditioning phase, second and fourth columns: post-conditioning phase; blue dots: rewarded, green dots: Null, and red dots:

unrewarded compartments. HIP activity formed a vane pattern, whilst NAc activity is rather polygonal with a polygonal pattern.

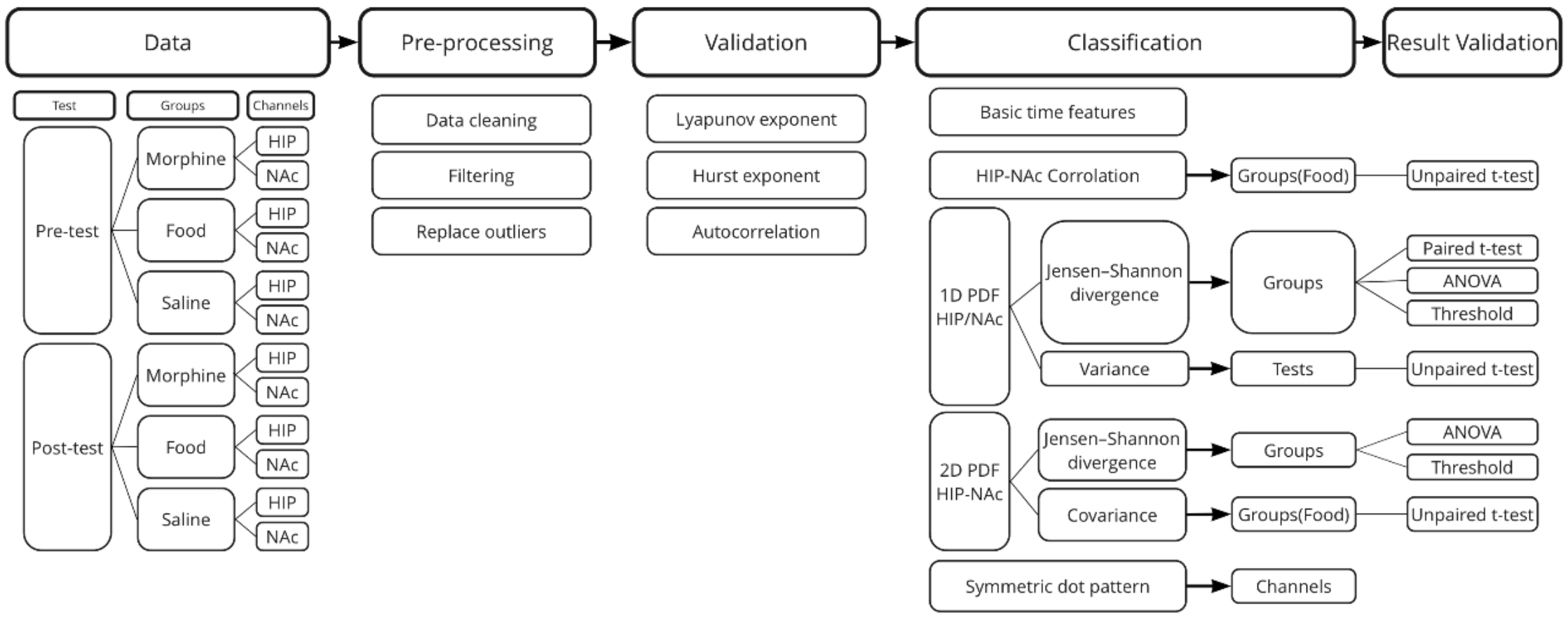

**Figure 4. The steps of classification**